# Second harmonic double resonance cones in dispersive hyperbolic metamaterials

Domenico de Ceglia,[1,*] Maria Antonietta Vincenti,[1] Salvatore Campione,[2] Filippo Capolino,[2] Joseph W. Haus[1,3], and Michael Scalora[4]

[1]*National Research Council – AMRDEC, Charles M. Bowden Research Laboratory, Redstone Arsenal, AL, 35898, USA*

[2]*Department of Electrical Engineering and Computer Science, University of California Irvine, CA, 92697, USA*

[3]*Electro-Optics Program, University of Dayton, 300 College Park, Dayton, OH, 45469, USA*

[4]*Charles M. Bowden Research Laboratory, AMRDEC, US Army RDECOM, Redstone Arsenal, AL, 35898, USA*

[*]*domenico.deceglia@us.army.mil*

**ABSTRACT**

We study the formation of second harmonic double-resonance cones in hyperbolic metamaterials. An electric dipole on the surface of the structure induces second harmonic light to propagate into two distinct volume plasmon-polariton channels: A signal that propagates within its own peculiar resonance cone; and a *phase-locked* signal that is trapped under the pump's resonance cone. Metamaterial dispersion and birefringence induce a large angular divergence between the two volume plasmon-polaritons, making these structures ideal for subwavelength second and higher harmonic imaging microscopy.

The inhomogeneous or *phase-locked* solution of nonlinear Helmholtz equation for harmonic generation processes [1, 2] travels at the same phase and group velocity of the fundamental frequency (FF) signal [3, 4]. It has also been shown that the generated phase-locked components survive in the presence of linear absorption at the harmonic frequencies [5]. In this Letter we predict that a hyperbolic metamaterial supports *resonance cones* (RCs), or *volume plasmon polaritons* (VPPs), in the presence of absorption at the fundamental and its harmonic frequencies, propagating in directions that are tilted with respect to the optical axis of the metamaterial.

Hyperbolic metamaterials can be implemented via layered metal-dielectric metamaterials. In such implementations, the study of harmonic generation should necessarily include material dispersion and absorption. Additionally, the artificial birefringence found in this class of metamaterials causes ordinary and extraordinary effective permittivities to have opposite signs, leading to a hyperbolic dispersion relation [6] and the propagation of large but finite spatial frequency components [7, 8]. In contrast, in the effective medium approximation (EMA) there is no upper bound to the maximum spectral component allowed in the medium. The further inclusion of cubic nonlinearities can improve the quality of subwavelength imaging by inhibiting diffraction via self-focusing, while maintaining broadband operation and increased propagation distances compared to linear metamaterials [9]. Favorable conditions for imaging through flat hyperbolic metamaterial lenses [10] are met when the ordinary effective permittivity $\varepsilon_\text{o}$ vanishes [11, 12], i.e., the permittivity component parallel to the metal-dielectric interfaces. Under these circumstances, the field distribution of an object at the input surface of the multilayer lens is imaged at the output plane with minimal distortion thanks to the formation of *parallel rays* that carry high spatial frequency components along the optical axis (*z*-axis, direction normal to the interfaces). Subwavelength imaging in metal-dielectric stacks can be further improved in the super-guiding [13, 14] and canalization [15] regimes, as described in Ref. [16] in terms of the point spread function. The multilayer focusing capabilities [13] may also be tailored by properly designing the curvature of the dispersion relation [17].

The chromatic dispersion of the metal permittivity limits the condition $\varepsilon_o \to 0$ to one specific wavelength, $\lambda_0$. Increasing the optical path length in the dielectric layers pushes $\lambda_0$ from the UV to the visible and infrared regions. For a given configuration, when $\lambda > \lambda_0$ the multilayer displays negative ordinary permittivity, whereas the extraordinary permittivity, i.e., the permittivity along the optical axis, is positive. In this regime, the stack still allows propagation of large spectral wave numbers, but not along the optical axis direction. Instead, the slab supports RCs [18] or VPPs [19] propagating in a *preferred direction* [20] tilted at some angle with respect to the optical axis.

Subwavelength interference of VPPs in a hyperbolic metamaterial substrate has been observed in a Young's double slit experiment [19]. It has been suggested that RCs may be used effectively to design single photon sources [21], but so far only the linear properties have been extensively investigated. Nevertheless, we will show that nonlinear optical interactions in these metamaterials may lead to exotic phenomena with potential for subwavelength imaging applications. We investigate the field patterns generated at the second harmonic (SH) frequency from electric dipole sources located on the surface of nonlinear, anisotropic slabs. While the FF signal diffracts in a RC with a single VPP, we predict that the diffraction of harmonic signals is channeled in two distinct VPPs: a *homogeneous* VPP that propagates in the RC of the harmonic frequency signal, and a *phase-locked* VPP, which is trapped in the RC of the FF signal.

We consider a hyperbolic metamaterial as in Fig. 1(b) made by alternating 5nm-thick, planar layers of silver and a generic dielectric medium having $\varepsilon_d=4$, compatible for example with metal-oxides like $Ta_2O_5$, $TiO_2$, $ZnO$, and with $LiNbO_3$ and $SrTiO_3$. The permittivity of silver $\varepsilon_m$ is taken from Palik's handbook [22]. The metal-dielectric interfaces lie on the *x-y* plane. By adopting the EMA, a simplified homogenization of this lamellar structure is based on a uniaxial, anisotropic model for the effective permittivity tensor $\boldsymbol{\varepsilon} = \varepsilon_o(\hat{\mathbf{x}}\hat{\mathbf{x}}+\hat{\mathbf{y}}\hat{\mathbf{y}})+\varepsilon_e\hat{\mathbf{z}}\hat{\mathbf{z}}$. In this model, ordinary and extraordinary permittivities, $\varepsilon_o = f\varepsilon_m + (1-f)\varepsilon_d$ and $\varepsilon_e = [f/\varepsilon_m + (1-f)/\varepsilon_d]^{-1}$ (in that order) depend on the metal fill factor $f = t_m/(t_m+t_d)$, where $t_m$ and $t_d$ are metal and dielectric layer thicknesses, respectively. We set $f = 0.5$. The problem of a line source radiating in a linear, planar [23] or cylindrical [24] metal-dielectric stack has been extensively studied. In the planar geometry, it is straightforward to demonstrate that the two-dimensional, scalar Green's function is singular when $(x-x')^2\varepsilon_e + (z-z')^2\varepsilon_o = 0$, where $(x,z)$ and $(x',z')$ are observation and source points, respectively. For a lossy, uniaxial crystal with $\text{Re}[\varepsilon_e]$ and $\text{Re}[\varepsilon_o]$ having opposite signs, the singularity manifests itself as a RC with a semi-angle

$$\theta_{RC} = \tan^{-1}\left(\sqrt{-\text{Re}(\varepsilon_o)/\text{Re}(\varepsilon_e)}\right), \quad (1)$$

evaluated with respect to the optical axis. In Fig. 1(a) we sum up the linear properties of the hyperbolic metamaterial described above by plotting the real parts of ordinary and extraordinary permittivities and the semi-angle of the RC, $\theta_{RC}$. The RC reduces to a single VPP propagating along the optical axis ($\theta_{RC}=0$) when the ordinary permittivity is zero, i.e. when $\lambda_0 = 405$nm. At larger wavelengths, VPPs propagate in RCs with angle $\theta_{RC} > 0$.

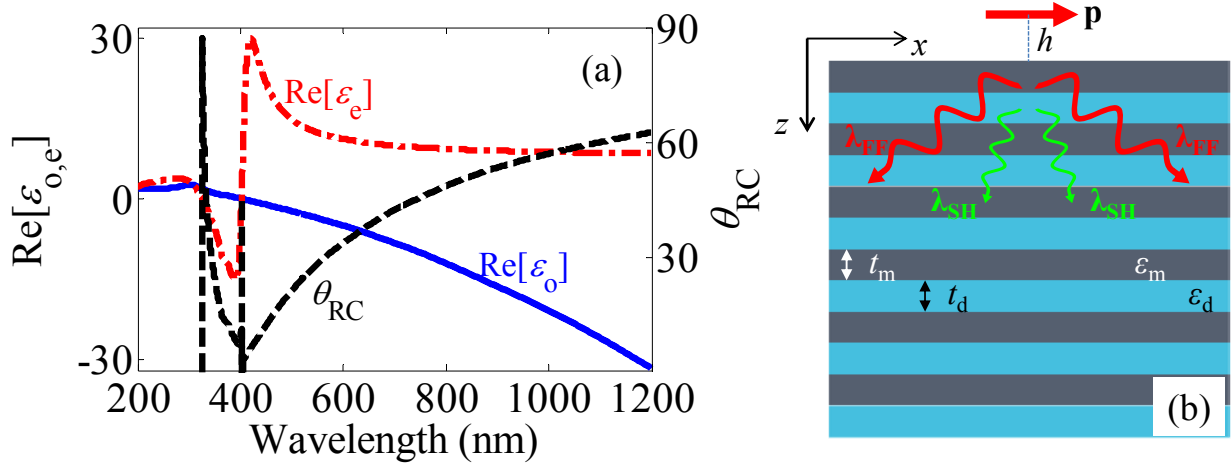

FIG. 1. (a) Real parts of the extraordinary (red, dash-dotted line) and ordinary (blue, solid line) permittivities of the uniaxial, anisotropic, effective medium model of a metal-dielectric stack with metal fill factor $f = 0.5$. The black, dashed line is the RC angle $\theta_{RC}$. (b) Schematic of an electric line dipole **p** on the surface of a hyperbolic metamaterial made of a semi-infinitely extended metal-dielectric multilayer. Red/green arrows represent FF/SH photons radiated into the structure.

We consider the field emitted by an electric line dipole $\mathbf{p} = p_x \hat{\mathbf{x}} e^{-i\omega t}$ located in free space at a distance $h = 5$ nm from a semi-infinite substrate shown in Fig. 1(b). The distance $h$ affects only the transverse size of VPPs without significantly altering RC angles. A detailed discussion on the effects of dipole-substrate distance is found in Refs. [8, 25]. We assume that the substrate, whose EMA properties are summarized in Fig. 1(a) has a homogeneous, instantaneous quadratic nonlinearity with non-zero susceptibility components $\chi^{(2)}_{xxx} = \chi^{(2)}_{zzz} = 1$ pm/V. The metamaterial scatters both FF and SH fields. Here we limit our attention to the fields radiated into the substrate region. We first tune the SH wavelength where $\theta_{RC} = 0$, so that $\lambda_{FF} = 2\lambda_{SH} = 2\lambda_0 = 810$ nm. For undepleted pumps, the FF signal propagates linearly, and the VPP at the FF is tilted at $\theta_{RC}(\lambda_{FF}) = 49°$ with respect to the $z$-axis − Fig. 1(a). The magnetic field intensity distribution $|H_y(x,z)|^2$ in the hyperbolic substrate at $\lambda_{FF} = 810$ nm − Fig. 2(a) − shows the propagation of subwavelength VPPs at $\theta_{RC}(\lambda_{FF}) = \pm 49°$.

The source of SH radiation is the electric current density $\mathbf{J}_{SH} = -i\omega_{SH}\varepsilon_0 \left( \chi^{(2)}_{xxx} E^2_{FF,x} \hat{\mathbf{x}} + \chi^{(2)}_{zzz} E^2_{FF,z} \hat{\mathbf{z}} \right)$ induced by the FF signal in the substrate region, where $\omega_{SH}$ is the SH angular frequency, $\varepsilon_0$ the free-space permittivity, and $E_{FF,x}$ and $E_{FF,z}$ are the FF electric field components in the $x$ and $z$ directions, respectively. Even if the intensity of the current $\mathbf{J}_{SH}$ peaks in the RC of the FF signal, i.e., at $\pm\theta_{RC}(\lambda_{FF})$, the SH field also shows maxima in the directions $\pm\theta_{RC}(\lambda_{SH})$, i.e., the RC at the SH wavelength. The generated SH radiation divides into two RCs: (i) The *phase-locked* SH cone at $\pm\theta_{RC}(\lambda_{FF})$ that overlaps with the RCs at the FF, and (ii) the *homogeneous* cone generated in the volume of the hyperbolic metamaterial just below the dipole. The SH component in the homogeneous cone walks off the RC at the FF, and freely propagates along the $\pm\theta_{RC}(\lambda_{SH})$ directions. For $\lambda_{FF} = 810$ nm, the *phase-locked* cone is tilted by $\theta_{RC}(\lambda_{FF}) \approx 49°$ with respect to the $z$ direction; the homogeneous cone $\pm\theta_{RC}(\lambda_{SH})$ is emitted at $\theta_{RC}(\lambda_{SH}) = 0°$, since $\lambda_{SH} = \lambda_0$. We note *that under TM-polarized light, a slab of stacked metal-dielectric layers with hyperbolic dispersion acts as a subwavelength imaging lens for SH light* [26], *thus providing strong spectral isolation from the excitation signal*.

In Fig. 2(c) and Fig. 2(d) we report the case for $\lambda_{FF} = 2\lambda_{SH} = 1010$ nm. Now, the FF and *phase-locked* SH cones are tilted by $\theta_{RC}(\lambda_{FF}) \approx 57°$; the *homogeneous* SH cone is found at $\theta_{RC}(\lambda_{SH}) \approx 22°$, in accordance with Fig. 1(a). These results show that the theory of homogeneous and *phase-locked* SH generation, applied earlier to plane waves or beams propagating in regimes of negligible diffraction [1-5], may be extended to subwavelength VPPs propagating in RCs of hyperbolic metamaterials. Waves with spectral wave numbers propagating along the *phase-locked* SH (or FF) cone travel at the same phase and group velocity, and attenuation rate of the FF. In contrast, waves with spectral wave numbers belonging to the *homogeneous* SH cone propagate with the phase and group velocity and attenuation rate of the hyperbolic medium at the SH frequency.

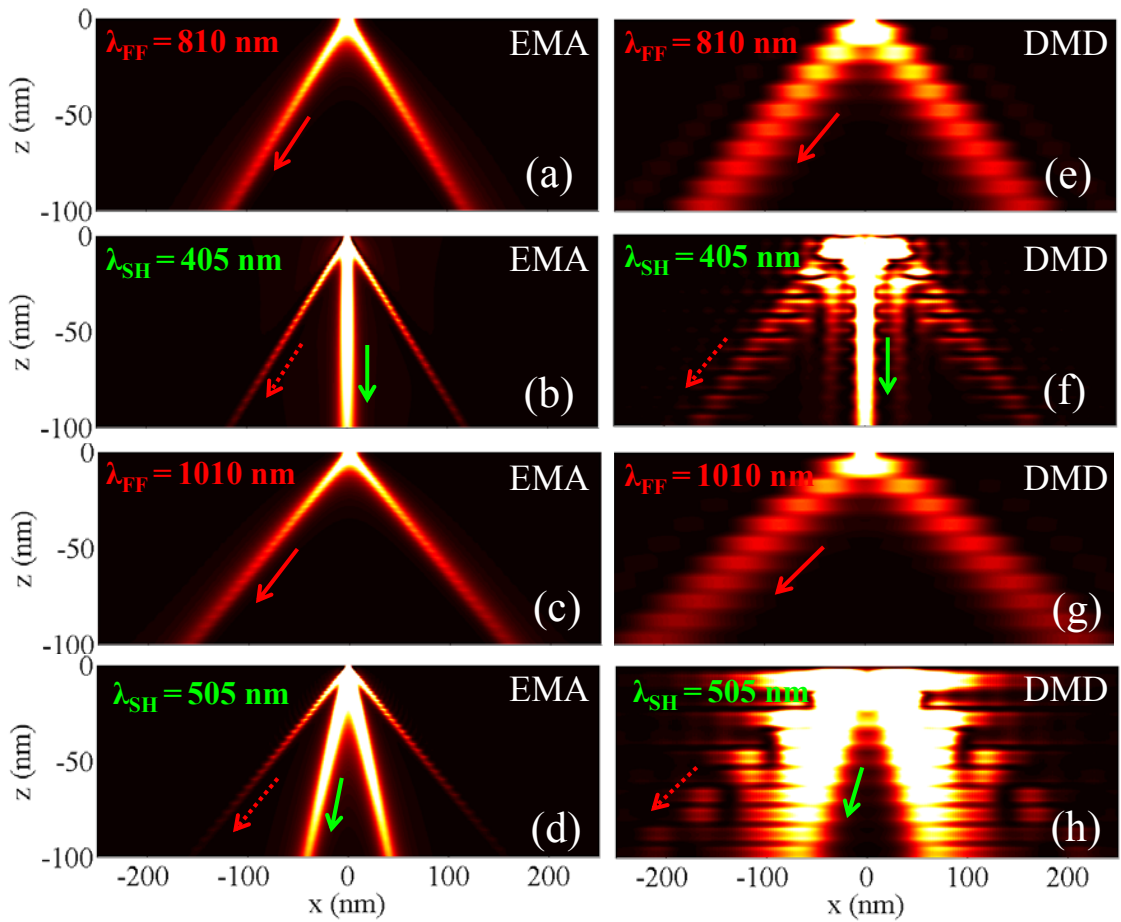

FIG. 2. (a) Magnetic field intensity distribution from a dipole located 5nm above the stack calculated with EMA. Dipole emission wavelength is $\lambda_{FF}$ = 810 nm. (b) Same as in (a), for the generated SH signal ($\lambda_{SH}$ = 405 nm). (c) Same as in (a), assuming $\lambda_{FF}$ = 1010 nm. (d) Same as in (b), at $\lambda_{SH}$ = 505 nm. (e), (f), (g) and (h), same as in (a), (b), (c), and (d), respectively, calculated for the DMD geometry. Solid and dashed red arrows point in the same direction since they are associated with the FF and phase-locked SH RCs, respectively. Green arrows indicate the homogeneous SH RC.

The anisotropic, effective medium model adopted above is valid strictly for metal-dielectric stacks with very small layers thickness. The homogenization model overstates the structure's ability to support propagation of large wave numbers. In fact, hyperbolicity of the dispersion relation is limited by layer thicknesses [7, 8, 25]. For these reasons we perform SH generation calculations by considering the discrete metal-dielectric (DMD) stack geometry − Fig. 1(b). We now assume the bulk nonlinearity is present only in the dielectric layers in the form: $\chi_{xxx}^{(2)} = \chi_{zzz}^{(2)} = 1$ pm/V. The magnetic field intensity distributions [$|H_y(x,z)|^2$] for the same FF/SH wavelengths as in Fig. 2(a-d) ($\lambda_{FF} = 2\lambda_{SH} = 2\lambda_0 = 810$ nm and $\lambda_{FF} = 2\lambda_{SH} = 1010$ nm) are now shown in Fig. 2 (e-h). The presence of VPPs propagating under RCs, and more importantly, the *phenomenology* of double-resonance-cone formation in the SH radiation pattern, are preserved when one compares the results from the EMA and those from the DMD geometry. Two RCs are clearly visible in the SH field distribution: homogeneous and phase-locked components. Despite the similarities, the models differ in several respects. For example, the RC angles are slightly larger and the transverse size of VPPs is wider for the DMD stack. In particular we find that, when $\lambda_{FF} = 2\lambda_{SH} = 2\lambda_0 = 810$ nm, FF and phase-locked SH RCs are found at 56° for the DMD case compared to 49° for the homogenized hyperbolic model, whereas the homogeneous RC is normal to the interfaces in both models. When $\lambda_{FF} = 2\lambda_{SH} = 1010$ nm, FF and phase-locked SH RCs are tilted at 62° for the DMD stack compared to 57° using the EMA, whereas the homogeneous RC is tilted by 34° for the DMD case compared to 22° in the homogenized hyperbolic model.

We now examine the differences between the EMA and the DMD stack. We gain insight into the physics of hyperbolic media using the Bloch theory for the bulk, plasmonic modes supported by the structure [8, 20]. In Figs. 3 (a) and 3(b) we plot the isofrequency diagrams of real and imaginary parts of the Bloch wave number for two wavelengths: (i) $2\lambda_0 = 810$ nm (red curves), and (ii) $\lambda_0 = 405$ nm (green curves), i.e., FF and SH wavelengths of the radiation patterns shown in Figs. 2(a) and 2(b) and Figs. 2(e) and 2(f). In the representation of Figs. 3(a) and 3(b), $k_x$ is the transverse wave number, assumed purely real and indicating each spectral component emitted by the dipole, $k_{B,z} = \beta_z + i\alpha_z$ is the complex, Bloch wave number, and $G = 2\pi / (t_m + t_d)$ is the magnitude of the reciprocal lattice vector. For comparison, isofrequency diagrams obtained using the homogenized, anisotropic dispersion relation $k_x^2/\varepsilon_e + k_z^2/\varepsilon_o = k^2$ are illustrated with dashed curves ($k$ is the free space wave number).

A significant deviation from the hyperbolic behavior predicted by the EMA is observed at 810 nm [i.e., the difference between solid and dashed red curves in Figs. 3 (a) and 3(b)]. The flattening of $\beta_z$ for high $k_x$ components and the associated increase of $\alpha_z$ are due to the modes' cutoff above the first Brillouin zone, in the region $\beta_z/G > 0.5$. The EMA is remarkably accurate at $\lambda_0 = 405$ nm, as suggested by the overlapping region of dashed and solid green curves in Figs. 3(a) and 3(b), and by looking at the virtually identical direction and transverse size of the homogeneous VPPs in the EMA [Fig. 2(b)] and in the DMD geometry [Fig. 2(f)].

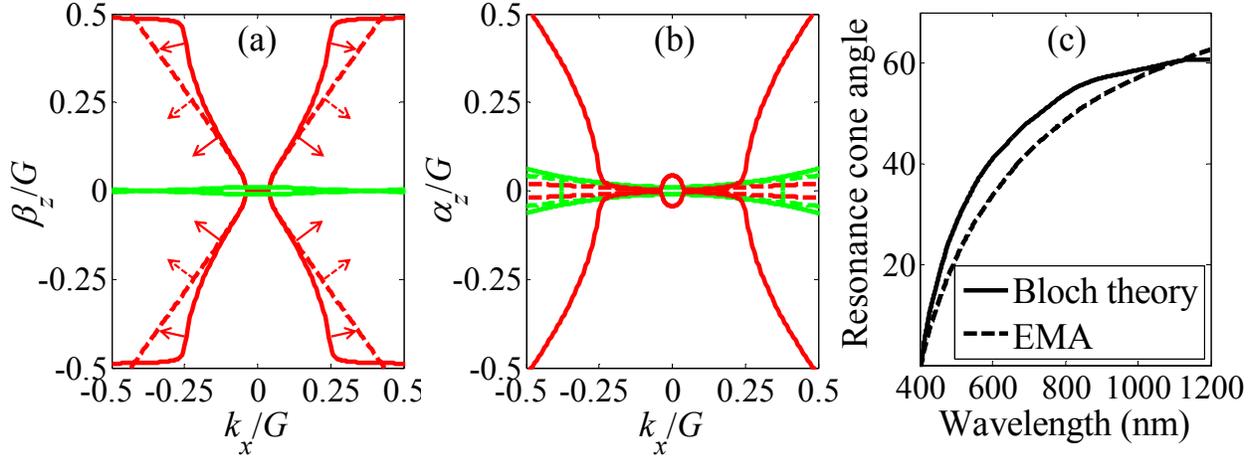

FIG. 3. Isofrequency diagrams at 810 nm (red) and 405 nm (green) in the $k_x$-$\beta_z$ (a) and $k_x$-$\alpha_z$ plane (b) for the metal-dielectric multilayer described in Fig. 1(b). Solid lines refer to Bloch theory, dashed lines to the anisotropic, hyperbolic model (EMA). Arrows indicate the direction of the group velocity; (c) RC angle with respect to the $z$-axis evaluated via Bloch theory, i.e., $\langle \theta_{BT} \rangle$ and with the anisotropic effective medium model, i.e., $\theta_{EMA}$.

Isofrequency diagrams also contain information about the direction of group velocity of each $k_x$-component. The group velocity, $\nabla_k \omega(\mathbf{k})$, is perpendicular to the frequency contours and points in the direction of increasing frequency $\omega$ [27]. This direction is evaluated as $\theta(k_x) = \tan^{-1}(\partial \beta_z / \partial k_x)$. We first analyze the diagrams related to the EMA [dashed curves in Figs. 3 (a) and 3(b)] and observe that the angle $\theta_{EMA}(k_x) = \tan^{-1}(\partial \beta_z / \partial k_x)$ is nearly constant. In other words, the dashed arrows in Fig. 3(a), which indicate $\theta_{EMA}(k_x)$ in the homogenized hyperbolic metamaterial, point in the same, *preferred* direction regardless of the chosen $k_x$-component. We note that this preferred direction coincides with the RC angle [Eq. (1)], derived by examining the shape of the Green's function singularity. This result is expected because both definitions of the RC angle, one in the spatial frequency domain [$\theta_{EMA}(k_x)$] based on the group velocity direction, the other in the real space domain ($\theta_{RC}$) based on the Green's function, originate from the EMA. Hence one may surmise that $\theta_{EMA}(k_x) \approx \theta_{RC}$. In contrast, the group velocity angle $\theta_{BT}(k_x) = \tan^{-1}(\partial \beta_z / \partial k_x)$ in the metal-dielectric multilayer modeled via Bloch theory is not constant. In fact, $\theta_{BT}(k_x)$ is very close to $\theta_{RC}$ only for small $k_x$-components ($|k_x|/G < 0.1$), becoming much larger than $\theta_{RC}$ for $k_x$-components near and above the edge of the first Brillouin zone ($|k_x|/G > 0.1$). This may be inferred by looking at the solid red arrows in Figs. 3(a), which indicate the group velocity direction in the DMD geometry, $\theta_{BT}(k_x)$. A zeroth order approximation of the RC angle in the DMD model is given by the weighted average

$$\langle \theta_{BT} \rangle = \int_0^\infty \alpha_z^{-1}(k_x) \theta_{BT}(k_x) dk_x \Big/ \int_0^\infty \alpha_z^{-1}(k_x) dk_x , \qquad (2)$$

where the function $\alpha_z^{-1}(k_x)$ weighs the radiation angle $\theta_{BT}(k_x)$ of each $k_x$-component with respect to its characteristic propagation distance (i.e., attenuation length) in the multilayer.

In Fig. 3(c) we show the wavelength dependence of the RC angle evaluated via the EMA and Bloch theory, defined by $\theta_{EMA}$ (dashed curve), and $\langle \theta_{BT} \rangle$ (solid curve), respectively. It turns out that in the wavelength range 405-1010nm, $\langle \theta_{BT} \rangle > \theta_{EMA}$. This result explains the discrepancies between radiation patterns evaluated with the two models at 810 nm [Figs. 2(a) and 2(e)] and the good agreement between the two models at $\lambda_0 = 405$ nm [Figs. 2(b) and 2(f)] where $\langle \theta_{BT} \rangle \approx \theta_{EMA}$.

In summary, we have discussed the formation of double RCs in dispersive hyperbolic metamaterials. We have demonstrated that a quadratic nonlinearity in anisotropic plasmas generates two subwavelength VPPs: One associated with the homogeneous SH component that propagates in a small-angle RC, the other *phase-locked* under the larger-angle RC at the FF. The differences between models based on EMA and DMD consist in a wider VPP cross section and larger RC angles in the metal-dielectric geometry, due to cutoff of high $k_x$-components near and above the first Brillouin zone. SH double RCs may drastically improve the subwavelength imaging abilities of metamaterial-based lenses thanks to the large spectral separation between the excitation and the observed signals. Double RCs are predicted in the radiation patterns at second and higher-order harmonic wavelengths in *any* anisotropic medium with hyperbolic dispersion. Although we have considered planar metal-dielectric multilayers, similar phenomenology is expected also in other implementations, such as nanowire metamaterials [28] and multilayer graphene structures [29, 30].

**REFERENCES**

1. J. A. Armstrong, N. Bloembergen, J. Ducuing, and P. S. Pershan, "Interactions between Light Waves in a Nonlinear Dielectric," Physical Review **127**, 1918-1939 (1962).

2. N. Bloembergen and P. S. Pershan, "Light Waves at the Boundary of Nonlinear Media," Physical Review **128**, 606-622 (1962).

3. L. D. Noordam, H. J. Bakker, M. P. de Boer, and H. B. v. L. van den Heuvell, "Second-harmonic generation of femtosecond pulses: observation of phase-mismatch effects: reply to comment," Opt. Lett. **16**, 971-971 (1991).

4. V. Roppo, M. Centini, C. Sibilia, M. Bertolotti, D. de Ceglia, M. Scalora, N. Akozbek, M. J. Bloemer, J. W. Haus, and O. G. Kosareva, "Role of phase matching in pulsed second-harmonic generation: Walk-off and phase-locked twin pulses in negative-index media," Physical Review A **76**, 033829 (2007).

5. M. Centini, V. Roppo, E. Fazio, F. Pettazzi, C. Sibilia, J. W. Haus, J. V. Foreman, N. Akozbek, M. J. Bloemer, and M. Scalora, "Inhibition of Linear Absorption in Opaque Materials Using Phase-Locked Harmonic Generation," Physical Review Letters **101**, 113905 (2008).

6. D. R. Smith, D. Schurig, J. J. Mock, P. Kolinko, and P. Rye, "Partial focusing of radiation by a slab of indefinite media," Applied Physics Letters **84**, 2244-2246 (2004).

7. O. Kidwai, S. V. Zhukovsky, and J. E. Sipe, "Dipole radiation near hyperbolic metamaterials: applicability of effective-medium approximation," Opt Lett **36**, 2530-2532 (2011).

8. C. Guclu, S. Campione, and F. Capolino, "Hyperbolic metamaterial as super absorber for scattered fields generated at its surface," Physical Review B **86**, 205130 (2012).

9. D. Aronovich and G. Bartal, "Nonlinear hyperlens," Opt. Lett. **38**, 413-415 (2013).

10. Z. Jacob, L. V. Alekseyev, and E. Narimanov, "Optical Hyperlens: Far-field imaging beyond the diffraction limit," Opt. Express **14**, 8247-8256 (2006).

11. S. A. Ramakrishna, J. B. Pendry, M. C. K. Wiltshire, and W. J. Stewart, "Imaging the near field," Journal of Modern Optics **50**, 1419-1430 (2003).

12. A. Salandrino and N. Engheta, "Far-field subdiffraction optical microscopy using metamaterial crystals: Theory and simulations," Physical Review B **74**, 075103 (2006).

13. M. Scalora, G. D'Aguanno, N. Mattiucci, M. J. Bloemer, D. de Ceglia, M. Centini, A. Mandatori, C. Sibilia, N. Akozbek, M. G. Cappeddu, M. Fowler, and J. W. Haus, "Negative refraction and sub-wavelength focusing in the visible range using transparent metallo-dielectric stacks," Opt. Express **15**, 508-523 (2007).

14. D. de Ceglia, M. Vincenti, M. Cappeddu, M. Centini, N. Akozbek, A. D'Orazio, J. Haus, M. Bloemer, and M. Scalora, "Tailoring metallodielectric structures for superresolution and superguiding applications in the visible and near-ir ranges," Physical Review A **77**, 033848 (2008).

15. P. A. Belov and Y. Hao, "Subwavelength imaging at optical frequencies using a transmission device formed by a periodic layered metal-dielectric structure operating in the canalization regime," Physical Review B **73**, 113110 (2006).

16. R. Kotyński and T. Stefaniuk, "Comparison of imaging with sub-wavelength resolution in the canalization and resonant tunnelling regimes," Journal of Optics A: Pure and Applied Optics **11**, 015001 (2009).

17. J. Bénédicto, E. Centeno, and A. Moreau, "Lens equation for flat lenses made with hyperbolic metamaterials," Opt. Lett. **37**, 4786-4788 (2012).

18. R. K. Fisher and R. W. Gould, "Resonance Cones in the Field Pattern of a Short Antenna in an Anisotropic Plasma," Physical Review Letters **22**, 1093-1095 (1969).

19. S. Ishii, A. V. Kildishev, E. Narimanov, V. M. Shalaev, and V. P. Drachev, "Sub-wavelength interference pattern from volume plasmon polaritons in a hyperbolic medium," Laser & Photonics Reviews **7**, 265-271 (2013).

20. B. Wood, J. B. Pendry, and D. P. Tsai, "Directed subwavelength imaging using a layered metal-dielectric system," Physical Review B **74**, 115116 (2006).

21. W. D. Newman, C. L. Cortes, and Z. Jacob, "Enhanced and directional single-photon emission in hyperbolic metamaterials," J. Opt. Soc. Am. B **30**, 766-775 (2013).

22. E. D. Palik and G. Ghosh, *Handbook of optical constants of solids* (Academic press, 1998), Vol. 3.

23. X. Ni, G. V. Naik, A. V. Kildishev, Y. Barnakov, A. Boltasseva, and V. M. Shalaev, "Effect of metallic and hyperbolic metamaterial surfaces on electric and magnetic dipole emission transitions," Applied Physics B **103**, 553-558 (2011).

24. H. Liu and K. J. Webb, "Resonance cones in cylindrically anisotropic metamaterials: a Green?s function analysis," Opt. Lett. **36**, 379-381 (2011).

25. O. Kidwai, S. V. Zhukovsky, and J. E. Sipe, "Effective-medium approach to planar multilayer hyperbolic metamaterials: Strengths and limitations," Physical Review A **85**, 053842 (2012).

26. A. V. Zayats and D. Richards, *Nano-optics and Near-field Optical Microscopy* (Artech House, Incorporated, 2009).

27. J. D. Joannopoulos, S. G. Johnson, J. N. Winn, and R. D. Meade, *Photonic crystals: molding the flow of light* (Princeton university press, 2011).

28. J. Elser, R. Wangberg, V. A. Podolskiy, and E. E. Narimanov, "Nanowire metamaterials with extreme optical anisotropy," Applied Physics Letters **89**, 261102-261103 (2006).

29. I. V. Iorsh, I. S. Mukhin, I. V. Shadrivov, P. A. Belov, and Y. S. Kivshar, "Hyperbolic metamaterials based on multilayer graphene structures," Physical Review B **87**, 075416 (2013).

30. M. A. K. Othman, C. Guclu, and F. Capolino, "Graphene-based tunable hyperbolic metamaterials and enhanced near-field absorption," Opt. Express **21**, 7614-7632 (2013).